\documentclass{amsart}
\usepackage{amsfonts}
\usepackage{amsmath}
\usepackage{amssymb}

\theoremstyle{plain}
\newtheorem{theorem}{Theorem}[section]

\newtheorem{proposition}[theorem]{Proposition}
\theoremstyle{definition}

\newtheorem{example}[theorem]{Example}

\theoremstyle{remark}
\newtheorem{remark}[theorem]{Remark}
\numberwithin{equation}{section}

\begin{document}
\title{The non-polynomial conservation laws and integrability analysis of
generalized Riemann type hydrodynamical equations}
\author{Ziemowit Popowicz}
\address{The Institute of Theoretical Physics, University of Wroc\l aw,
Poland}
\email{ziemek@ift.uni.wroc.pl}
\author{Anatoliy K. Prykarpatsky}
\address{ The AGH University of Science and Technology, Krakow 30059,
Poland, and the Ivan Franko State Pedagogical University, Drohobych, Lviv
region, Ukraine}
\email{pryk.anat@ua.fm}

\begin{abstract}
Based on the gradient-holonomic algorithm we analyze the integrability
property of the generalized hydrodynamical Riemann type equation $%
D_{t}^{N}u=0$ for arbitrary $N\in \mathbb{Z}_{+}.$ The infinite hierarchies
of polynomial and non-polynomial conservation laws, both dispersive and
dispersionless are constructed. Special attention is paid to the cases $%
N=2,3 $ and $N=4,\ $\ for which the conservation laws, Lax type
representations and bi-Hamiltonian structures are analyzed in detail. We
also show that the case $N=2$ is equivalent to a generalized Hunter-Saxton
dynamical system, whose integrability follows from the results obtained. As
a byproduct of our analysis we demonstrate a new set of non-polynomial
conservation laws for the related Hunter-Saxton equation.
\end{abstract}

\maketitle

\section{\protect\bigskip Introduction}

Nonlinear hydrodynamic equations are of constant interest still from
classical works by B. Riemann, who had extensively studied them in general
three-dimensional case, having paid special attention to their
one-dimensional spatial reduction, for which he devised the generalized
method of characteristics and Riemann invariants. \ These methods appeared
to be very effective \cite{Wh,PBB,PM} in investigating many types of
nonlinear spatially one-dimensional systems of hydrodynamical type and, in
particular, the characteristics method in the form of a "reciprocal"
transformation of variables has been \ used recently in studying a so called
Gurevich-Zybin system \cite{GZ,GZ1} in \cite{Pav} and a Whitham type system
in \cite{PM,Sak}. Moreover, this method was further effectively applied to
studying solutions to a generalized \cite{GPPP} (owing to D. Holm and M.
Pavlov) Riemann type hydrodynamical system
\begin{equation}
D_{t}^{N}u=0,\text{ \ \ }D_{t}:=\partial /\partial t+u\partial /\partial x,%
\text{ \ }N\in \mathbb{Z}_{+},  \label{H0a}
\end{equation}%
where $dx/dt=u\in C^{\infty }(\mathbb{R};\mathbb{R})$ is the corresponding
characteristic flow velocity along the real axis $\mathbb{R}.$

We will consider, for convenience, the hydrodynamical equation (\ref{H0a})
on the $2\pi $-periodic space of functions $\mathcal{M}_{0}:=C^{\infty }(%
\mathbb{R}/2\pi \mathbb{Z};\mathbb{R}),$ \ which can be, obviously,
equivalently rewritten as the following nonlinear dynamical system in the
augmented functional manifold $\mathcal{M}:=C^{\infty }(\mathbb{R}/2\pi
\mathbb{Z};\mathbb{R}^{N})$ on the vector $\ \hat{u}\mathbf{:=(}u^{(0)}:=u,$
$u^{(1)}:=D_{t}u^{(0)},$ $u^{(2)}:=$ $D_{t}u^{(1)},...,u^{(N-1)}$ $%
:=D_{t}u^{(N-2)})^{\intercal }$ $\in \mathcal{M}:$%
\begin{eqnarray}
u_{t}^{(0)} &=&u^{(1)}-u^{(0)}u_{x}^{(1)},  \notag \\
u_{t}^{(1)} &=&u^{(2)}-u^{(0)}u_{x}^{(2)},  \notag \\
&&......  \label{H0b} \\
u_{t}^{(N-2)} &=&u^{(N-1)}-u^{(0)}u_{x}^{(N-2)}  \notag \\
u_{t}^{(N-1)} &=&-u^{(1)}u_{x}^{(N-1)}.  \notag
\end{eqnarray}%

The dynamical system (\ref{H0b}) possesses a very interesting and important
property: the partial flows of the velocity components $u^{(j)},j=\overline{%
0,N-1},$ are realized along the axis $\mathbb{R}$ with the same
characteristic velocity $dx/dt=u^{(0)}.$ This is exactly the case, deeply
studied by Riemann (see, for example, \cite{Wh}) when one can introduce the
so-called "Riemann invariants" making it possible \ to obtain a suitable
separation of dependent variables important for the integration. Really, we
can observe that the system (\ref{H0b}) is equivalent along the
characteristics $dx/dt=u^{(0)}$\ to the following recurrent set of
differential equations in full differentials:
\begin{eqnarray}
du^{(0)} &=&u^{(1)}dt,  \notag \\
du^{(1)} &=&u^{(2)}dt,  \notag \\
&&......  \label{H0c} \\
du^{(N-2)} &=&u^{(N-1)}dt,  \notag \\
du^{(N-1)} &=&0,  \notag
\end{eqnarray}%
whose solution is easily found by means of simple integration in the
parametric form as

\begin{eqnarray}
u^{(N-1)} &:&=z,\text{ \ }z=\text{\ }\beta _{N}(x-\frac{zt^{N}}{N!}%
-\sum_{j=1}^{N-1}\frac{t^{j}}{j!}\beta _{N-j}(z)),  \notag \\
u^{(N-2)} &=&zt+\beta _{1}(z),\text{ \ \ }u^{(N-3)}=\frac{zt^{2}}{2!}+t\beta
_{1}(z)+\beta _{2}(z),  \notag \\
u^{(N-3)} &=&\frac{zt^{3}}{3!}+\frac{t^{2}}{2!}\beta _{1}(z),  \notag \\
&&......  \label{H0d} \\
u^{(1)} &=&\frac{zt^{N-2}}{(N-2)!}+\sum_{j=0}^{N-3}\frac{t^{j}}{j!}\beta
_{N-2-j}(z),  \notag \\
u^{(0)} &=&\frac{zt^{N-1}}{(N-1)!}+\sum_{j=0}^{N-2}\frac{t^{j}}{j!}\beta
_{N-1-j}(z),\text{ \ \ \ \ }  \notag
\end{eqnarray}%
where $\beta _{j}\in C^{\infty }(\mathbb{R};\mathbb{R}),j=\overline{1,N-1},$
and $\beta _{N}\in C^{\infty }(\mathbb{R}/2\pi \mathbb{Z};\mathbb{R})$ \ are
arbitrary smooth functions, depending on a suitable first integral $z\in
C^{\infty }(\mathbb{R}^{N};\mathbb{R})$ of the system (\ref{H0c}). The
presented above result coincides in some part with that obtained before by
M. Pavlov in \cite{BGPPP}, and can be used for constructing special
solutions in analytical form to the generalized Riemann type hydrodynamical
equation (\ref{H0a}).

As it was stated before in \cite{PP,Pav,DB,PrPryt,BGPPP,GPPP} the Riemann
type hydrodynamical system (\ref{H0b}) at $N=2$ and $N=3$ possesses
additional very interesting properties, being a completely integrable
bi-Hamiltonian system. In particular, it possesses infinite hierarchies of
dispersionless and dispersive conservation laws, which can have an important
hydrodynamical interpretation and may be used for constructing a wide class
of other special quasi-periodic and solitonic solutions.

In spite of the exact integrability of dynamical system (\ref{H0a}) by means
of the classical characteristics method, the solutions obtained this way
are, to the regret, of very vague usefulness, as they are given in the
entangled and involved form not fitting for studying solutions belonging to
some specially assigned classes of functions, for instance, fast-decreasing,
quasi-periodic and etc. Thereby, additional studying of the mathematical
structures associated with dynamical system (\ref{H0a}) by means of modern
symplectic theory techniques is as much as could needed, and what is a topic
of our present investigation.

The next section below is devoted to the Hamiltonian analysis of the
hydrodynamical system (\ref{H0b}) at $N=2,$ $N=3$ and $N=4,$ as well as to
the description of their new hierarchies of conservation laws, the related
co-symplectic structures and Lax type representations.

\bigskip\ \

\section{The generalized Riemann type hydrodynamical equation at N=2:
conservation laws, bi-Hamiltonian structure and Lax type representation}

Consider the generalized Riemann type hydrodynamical equation (\ref{H0a}) at
$N=2:$
\begin{equation}
D_{t}^{2}u=0,  \label{H0}
\end{equation}%
where $D_{t}=\partial /\partial t+u\partial /\partial x,$ which is
equivalent to the following dynamical system:%
\begin{equation}
\left.
\begin{array}{c}
u_{t}=v-uu_{x} \\
v_{t}=-uv_{x}%
\end{array}%
\right\} :=K[u,v],  \label{H1-0}
\end{equation}%
where $K:\mathcal{M}\rightarrow T(\mathcal{M)}$ \ is a related vector field
on the $2\pi $-periodic smooth nonsingular functional phase space $\mathcal{M%
}:=\{(u,v)^{\intercal }$ $\in C^{\infty }(\mathbb{R}/2\mathbb{\pi Z};\mathbb{%
R}^{2})$ $:u_{x}^{2}-2v_{x}\neq 0,x\in \mathbb{R}\}.$ As we are interested
first in the conservation laws for the system (\ref{H1-0}), the following
proposition holds.

\begin{proposition}
\label{Pr_1} Let $H(\lambda ):=\int_{0}^{2\pi }h(x;\lambda )dx\in D(\mathcal{%
M})$ be an almost everywhere smooth functional on the manifold $\mathcal{M},$
depending parametrically on $\lambda \in \mathbb{C},$ and whose density
satisfies the differential condition%
\begin{equation}
h_{t}=\lambda (uh)_{x}  \label{H1}
\end{equation}%
for all $t\in \mathbb{R}$ and $\lambda \in \mathbb{C}$ on the solution set
of dynamical system (\ref{H1-0}). Then the \ following iterative
differential relationship
\begin{equation}
(f/h)_{t}=\lambda (uf/h)_{x}  \label{H2}
\end{equation}%
holds, if a smooth function $f\in C^{\infty }(\mathbb{R};\mathbb{R})$
(parametrically depending on $\lambda \in \mathbb{C)}$ satisfies for all $%
t\in \mathbb{R}$ the linear equation
\begin{equation}
f_{t}=2\lambda u_{x}f+\lambda uf_{x}.  \label{H3}
\end{equation}
\end{proposition}

\begin{proof}
We have from (\ref{H1})-(\ref{H3}) that
\begin{eqnarray}
(f/h)_{t} &=&f_{t}/h-fh_{t}/h^{2}=f_{t}/h-\lambda fu_{x}/h-\lambda
fuh_{x}/h^{2}=  \label{H4} \\
&=&f_{t}/h+\lambda fu(1/h)_{x}-\lambda u_{x}f/h=  \notag \\
&=&\lambda (uf)_{x}/h+\lambda uf(1/h)_{x}=\lambda (uf/h)_{x},  \notag
\end{eqnarray}%
proving the proposition.
\end{proof}

The obvious generalization of the previous proposition is read as follows.

\begin{proposition}
\label{Pr_2}If a smooth function $h\in C^{\infty }(\mathbb{R};\mathbb{R})$
satisfies the relationship
\begin{equation}
h_{t}=ku_{x}h+uh_{x},  \label{H1a}
\end{equation}%
where $k\in \mathbb{R},$ then
\begin{equation}
H=\int_{0}^{2\pi }h^{1/k}dx  \label{H1b}
\end{equation}%
is a conservation law for the Riemann type hydrodynamical system (\ref{H1-0}%
).
\end{proposition}

\begin{remark}
\label{Rm_1}Let $\hat{h}$ $\in C^{\infty }(\mathbb{R};\mathbb{R})$ satisfy
the differential relationship $\hat{h}_{t}=(\hat{h}u)_{x},$ then $f=\hat{h}%
^{2}$ is a solution to equation (\ref{H2}).
\end{remark}

\begin{remark}
\label{Rm_2}If functions $h_{j}\in C^{\infty }(\mathbb{R};\mathbb{R}),j\in
\mathbb{Z}_{+},$ satisfy the relationships $h_{j,t}=\lambda
(h_{j}u)_{x},j\in \mathbb{Z}_{+},$ $\lambda \in \mathbb{C},$ then the
functionals
\begin{equation}
H_{(i,j)}=\sum_{n\in \text{ }\mathbb{Z}_{+}\mathbb{\ }}k_{n}^{(i,j)}%
\int_{0}^{2\pi }h_{i}^{2^{n}}h_{j}^{(1-2^{n})}  \label{H1c}
\end{equation}%
with $k_{n}^{(i,j)}$ $\in \mathbb{R},$ $\ n\in \mathbb{Z}_{+},$ $i,j\in
\mathbb{Z}_{+},$ being arbitrary constants, are conserved quantities to
equation (\ref{H1-0}). This formula, in particular, makes it possible to
construct an infinite hierarchy of non-polynomial conserved quantities for
the Riemann type hydrodynamical system (\ref{H1-0}).
\end{remark}


\begin{example}
The following non-polynomial functionals
\begin{eqnarray}
H_{4}^{(\frac{1}{3})} &=&\int_{0}^{2\pi }\sqrt{u_{x}^{2}-2v_{x}}dx,\quad
\quad H_{7}^{(\frac{1}{3})}=\int_{0}^{2\pi }\left(
u_{x}v_{xx}-u_{xx}v_{x}\right) ^{1/3}dx,  \notag  \label{hamy} \\
&&  \notag \\
\text{ \ }H_{7}^{(\frac{1}{2})} &=&\int_{0}^{2\pi }\sqrt{v(u_{x}^{2}-2v_{x})}%
dx,  \notag \\
&&  \notag \\
H_{8}^{(\frac{1}{3})} &=&\int_{0}^{2\pi }\left(
k_{1}u(u_{xx}v_{x}-u_{x}v_{xx})+k_{1}v_{xx}v+k_{2}(u_{x}^{2}v-2v_{x}^{2})%
\right) ^{1/3}dx,  \notag \\
&&  \label{H5} \\
H_{9}^{(\frac{1}{6})} &=&\int_{0}^{2\pi }\left(
u_{xx}v_{xxx}-u_{xxx}v_{xx}\right) ^{\frac{1}{6}}dx,\quad  \notag \\
&&  \notag \\
H_{9}^{(\frac{1}{4})} &=&\int_{0}^{2\pi }\left(
u_{x}(u_{xx}v_{x}-u_{x}v_{xx})+v_{xx}v_{x})\right) ^{\frac{1}{4}},  \notag \\
&&  \notag \\
H_{10}^{(\frac{1}{6})} &=&\int_{0}^{2\pi }\left(
2u_{xx}(u_{x}v_{xx}-u_{xx}v_{x})-v_{xx}^{2}\right) ^{\frac{1}{6}}  \notag
\end{eqnarray}

are conservation laws for the Riemann type dynamical system (\ref{H1-0}).
\end{example}

Quite different conservation laws have been obtained in \cite{DB,Pav} using
the recursion operator technique. The corresponding recursion operator
proves to generate no new conservation law, if one applies it to the
non-polynomial conservations laws (\ref{H5}).

We also notice that dynamical system (\ref{H1-0}), as it was shown before in
\cite{Pav,BPrGG}, can be transformed via the substitution
\begin{equation}
v=\frac{1}{2}\partial ^{-1}(u_{x}^{2}+\eta ^{2})  \label{H6}
\end{equation}%
into the generalized two-component Hunter - Saxton equation:%
\begin{eqnarray}
u_{x,t} &=&-\frac{1}{2}u_{x}^{2}-uu_{xx}+\frac{1}{2}\eta ^{2},  \label{H7} \\
\eta _{t} &=&-(u\eta )_{x}.  \notag
\end{eqnarray}%
This equation allows the simple reduction to the Hunter - Saxton dynamical
system \cite{HS,PrPryt,BPrGG} at $\eta =0:$%
\begin{equation}
u_{xt}=-\frac{1}{2}u_{x}^{2}-uu_{xx}.  \label{H7a}
\end{equation}%
The non-polynomial conservation laws (\ref{H5}), upon rewriting with respect
to the substitution (\ref{H6}), give rise to the related non-polynomial
conservations laws for the generalized two-component Hunter - Saxton
dynamical system (\ref{H7}). Moreover, if we further apply the reduction $%
\eta =0,$ we obtain, respectively, new non-polynomial conservation laws for
the Hunter - Saxton dynamical system (\ref{H7a}), supplementing those found
before in \cite{PrPryt,BPrGG}.

\begin{example}
The following functionals
\begin{eqnarray}
H_{7}^{(\frac{1}{3})} &=&\int_{0}^{2\pi }\big (u_{xx}u_{x}^{2}\big )^{\frac{1%
}{3}}dx,\text{ \ }H_{9}^{(\frac{1}{6})}=\int_{0}^{2\pi }\frac{%
u_{xxx}u+2u_{xx}u_{x}}{\sqrt{u_{xx}}}dx,  \label{H5a} \\
H_{8}^{(\frac{1}{3})} &=&\int_{0}^{2\pi }\big [u_{xx}u_{x}(\partial
^{-1}u_{x}^{2})-u_{xx}u_{x}^{2}u\big ]^{\frac{1}{3}}dx  \notag
\end{eqnarray}%
are the conservation laws for the Hunter-Saxton dynamical system (\ref{H7a}).
\end{example}

All of these and many others non-polynomial conservation laws can be easily
obtained using Proposition (\ref{Pr_2}). \ For example, the next functionals

\begin{eqnarray}
H_{(n,m)} &=&\int_{0}^{2\pi }\big (u_{xx}^{n}u_{x}^{m}\big )^{\frac{2}{m+4n}%
}dx,\text{ \ \ }H_{(1)}=\int_{0}^{2\pi }u_{x}^{2}(\partial
^{-1}u_{x}^{2})^{2}dx,  \notag \\
H_{(2)} &=&\int_{0}^{2\pi }\sqrt{u_{xx}}dx,\text{ \ \ \ \ }%
H_{(3)}=\int_{0}^{2\pi }\sqrt{u_{xx}(\partial ^{-1}u_{x}^{2})}dx,
\label{H5b} \\
H_{(4)} &=&\int_{0}^{2\pi }\big [(\partial
^{-1}u_{x}^{2})(uu_{x}u_{xx}^{2}-u_{xx}^{2}(\partial ^{-1}u_{x}^{2}))\big ]^{%
\frac{2}{9}}dx  \notag
\end{eqnarray}%
are also conservation laws for the Hunter-Saxton dynamical system (\ref{H7a}%
), where $m\neq -4n\ $\ and $\ n,m\in \mathbb{Z}.$

Now we proceed to analyzing the Hamiltonian properties of the dynamical
system (\ref{H1-0}), for which we will search for solutions to the
determining \cite{PM,Bl} N\"{o}ther equation
\begin{equation}
L_{K}\vartheta =\vartheta _{t}-\vartheta K^{\prime ,\ast }-K^{\prime
}\vartheta =0.  \label{H8}
\end{equation}%
where $L_{K}$ denotes the corresponding Lie derivative on $\mathcal{M}$
subject to $\ $the vector field $K:\mathcal{M}\rightarrow T(\mathcal{M)},$ \
$K^{\prime }:T(\mathcal{M)\rightarrow }T(\mathcal{M)}$ is its Frechet
derivative, $K^{\prime \ast }:T^{\ast }(\mathcal{M)\rightarrow }T^{\ast }(%
\mathcal{M)}$ is its conjugation with respect to the standard bilinear form $%
(\cdot ,\cdot )$ on $T^{\ast }(\mathcal{M)\times }T(\mathcal{M)},$ and $%
\vartheta :T^{\ast }(\mathcal{M})\rightarrow T(\mathcal{M})$ is a suitable
implectic operator on $\mathcal{M},$ with respect to which the following
Hamiltonian representation
\begin{equation}
K=-\vartheta \text{ }grad\text{ }H_{\vartheta }  \label{H9}
\end{equation}%
for some smooth functional $H_{\vartheta }\in D(\mathcal{M})$ holds. To show
this, it is enough to find, for instance by means of the small parameter
method \cite{PM,HPP}, a non-symmetric $(\psi ^{\prime }\neq \psi ^{\prime
,\ast })$ solution $\psi \in T^{\ast }(\mathcal{M})$ to the following
Lie-Lax equation:%
\begin{equation}
\psi _{t}+K^{\prime ,\ast }\psi =grad\text{ }\mathcal{L}  \label{H10}
\end{equation}%
for some suitably chosen smooth functional $\mathcal{L}\in \mathcal{D}(M).$
As a result of easy calculations one obtains that
\begin{equation}
\psi =(v,0)^{\intercal },\text{ \ \ }L=\frac{1}{2}\int_{0}^{2\pi }v^{2}dx.
\label{H11}
\end{equation}%
Making use of (\ref{H10}) jointly with the classical Legendrian relationship
\begin{equation}
H_{\vartheta }:=(\psi ,K)-\mathcal{L}  \label{H12}
\end{equation}%
for the suitable Hamiltonian function, one easily obtains the corresponding
symplectic structure
\begin{equation}
\vartheta ^{-1}:=\psi ^{\prime }-\psi ^{\prime ,\ast }=\left(
\begin{array}{cc}
0 & 1 \\
-1 & 0%
\end{array}%
\right)  \label{H13}
\end{equation}%
and the non-singular Hamilton function
\begin{equation}
H_{\vartheta }:=\frac{1}{2}\int_{0}^{2\pi }(v^{2}+v_{x}u^{2})dx.  \label{H14}
\end{equation}%
Since the operator (\ref{H13}) is nonsingular, we obtain the corresponding
implectic operator
\begin{equation}
\vartheta =\left(
\begin{array}{cc}
0 & -1 \\
1 & 0%
\end{array}%
\right) ,  \label{H13a}
\end{equation}%
necessarily satisfying the N\"{o}ther equation (\ref{H8}).

It is worth here to observe that the determining Lie-Lax equation (\ref{H10}%
) possesses still other solution
\begin{equation}
\psi =(\frac{u_{x}}{2},-\frac{u_{x}^{2}}{2v_{x}}),\text{ }\mathcal{L}=\frac{1%
}{4}\int_{0}^{2\pi }uv_{x}dx,  \label{H15}
\end{equation}%
giving rise, owing to expressions (\ref{H13}) and (\ref{H12}), to the new
co-implectic (singular "symplectic") structure
\begin{equation}
\eta ^{-1}:=\psi ^{\prime }-\psi ^{\prime ,\ast }=\left(
\begin{array}{cc}
\partial & -\partial u_{x}v_{x}^{-1} \\
-u_{x}v_{x}^{-1}\partial & u_{x}^{2}v_{x}^{-2}\partial +\partial
u_{x}^{2}v_{x}^{-2}%
\end{array}%
\right)  \label{H16}
\end{equation}%
on the manifold $\mathcal{M},$ subject to which the Hamiltonian functional
equals%
\begin{equation}
H_{\eta }:=\frac{1}{2}\int_{0}^{2\pi }(u_{x}v-v_{x}u)dx,  \label{H17}
\end{equation}%
supplying the second Hamiltonian representation
\begin{equation}
K=-\eta \text{ }{grad}\text{ }H_{\eta }  \label{H9b}
\end{equation}%
of the Riemann type hydrodynamical system (\ref{H1-0}). The co-implectic
structure (\ref{H16}) is singular, since $\hat{\eta}^{-1}(u_{x},v_{x})^{%
\intercal }=0,$ nonetheless one can calculate its inverse expression
\begin{equation}
\eta :=\left(
\begin{array}{cc}
-\partial ^{-1} & u_{x}\partial ^{-1} \\
\partial ^{-1}u_{x} & v_{x}\partial ^{-1}+\partial ^{-1}v_{x}%
\end{array}%
\right) ,  \label{H18}
\end{equation}%
Moreover, the corresponding implectic structure $\eta :T^{\ast }(\mathcal{M}%
)\rightarrow T^{\ast }(\mathcal{M})$ satisfies the determining N\"{o}ther
equation
\begin{equation}
L_{K}\eta =\eta _{t}-\eta K^{\prime ,\ast }-K^{\prime }\eta =0,  \label{H19}
\end{equation}%
whose solutions can also be obtained by means of the small parameter method
\cite{PM,MBPS}. \ We remark also that, owing to the general symplectic
theory results \cite{PM,MBPS,HPP,Bl} for nonlinear dynamical systems on
smooth functional manifolds, operator (\ref{H16}) defines on the manifold $%
\mathcal{M}$ a closed functional differential two-form. Thereby it is
\textit{a priori} co-implectic (in general, singular symplectic), satisfying
on $\mathcal{M}$ the standard Jacobi brackets condition.

As a result, the second implectic operator (\ref{H18}), being compatible
\cite{PM,Bl} with the implectic operator (\ref{H13a}), gives rise to a new
infinite hierarchy of polynomial conservation laws

\begin{equation}
\gamma _{n}:=\int_{0}^{1}d\lambda <(\vartheta ^{-1}\eta )^{n}{grad}\text{ }%
H_{\vartheta }[u\lambda \},u>  \label{H20}
\end{equation}%
for all $n\in \mathbb{Z}_{+}.$ Having defined the recursion operator $%
\Lambda :=\vartheta ^{-1}\eta ,$ one also finds from (\ref{H20}), (\ref{H8})
and (\ref{H19}) that the following Lax type relationship
\begin{equation}
L_{K}\Lambda =\Lambda _{t}-[\Lambda ,K^{^{\prime },\ast }]=0  \label{H21}
\end{equation}%
holds. If to construct now the asymptotical expansion $\varphi (x;\lambda
)\simeq \sum\limits_{j\in \mathbb{Z}_{+}}\lambda ^{1-2j}{grad}$ $\gamma
_{j-1}[u,v]$ as $\lambda \rightarrow \infty ,$ it is easy to obtain from (%
\ref{H20}) that the gradient like relationship%
\begin{equation}
\lambda ^{2}\vartheta \varphi (x;\lambda =\eta \varphi (x;\lambda )
\label{H21a}
\end{equation}%
holds. The latter relationship, making use of the implectic operators (\ref%
{H13a}) and (\ref{H18}), can be represented in the following two factorized
forms:
\begin{equation}
\varphi (x;\lambda ):=\left(
\begin{array}{c}
\varphi _{1}(x;\lambda ) \\
\varphi _{2}(x;\lambda )%
\end{array}%
\right) =\left(
\begin{array}{c}
-4\lambda ^{3}f_{1}^{2}+2\lambda v_{x}f_{2}^{2} \\
-4\lambda ^{2}f_{1}f_{2}-2\lambda u_{x}f_{2}^{2}%
\end{array}%
\right) =\left(
\begin{array}{c}
-2\lambda (f_{1}f_{2})_{x} \\
-(f_{2}^{2})_{x}%
\end{array}%
\right) ,  \label{H21b}
\end{equation}%
where a vector $f\in C^{\infty }(\mathbb{R}^{2};\mathbb{C}^{2})$ lies in an
associated to manifold $\mathcal{M}$ vector bundle $\mathcal{L(M};\mathbb{E}%
^{2}),$ whose fibers are isomorphic to the complex Euclidean vector space $%
\mathbb{E}^{2}.$ Take now into account \cite{PM,HPP} that the Lie-Lax
equation
\begin{equation}
L_{K}\varphi (x;\lambda )=d\varphi (x;\lambda )/dt+K^{\prime ,\ast }\varphi
(x;\lambda )=0  \label{H22}
\end{equation}%
can be transformed equivalently for all $x,t\in \mathbb{R}$ and $\lambda \in
\mathbb{C}$ into the following evolution system:
\begin{equation}
D_{t}\varphi =\left(
\begin{array}{cc}
0 & v_{x} \\
-1 & -u_{x}%
\end{array}%
\right) \varphi ,\text{ \ }D_{t}=\partial /\partial t+u\partial /\partial x.
\label{H22a}
\end{equation}%
The equation (\ref{H22a}), owing to the relationship (\ref{H21a}) and the
obvious identity
\begin{equation}
D_{t}f_{x}+u_{x}f_{x}=(D_{t}f)_{x},  \label{H21c}
\end{equation}%
can be further split into the adjoint to (\ref{H22a}) system
\begin{equation}
D_{t}f=q(\lambda )f,\text{ \ \ }q(\lambda ):=\left(
\begin{array}{cc}
0 & 0 \\
-\lambda & 0%
\end{array}%
\right) ,\text{ \ \ }  \label{H23}
\end{equation}%
where a vector $f\in C^{\infty }(\mathbb{R}^{2};\mathbb{C}^{2})$ satisfies
the following linear equation%
\begin{equation}
f_{x}=\ell \lbrack u,v;\lambda ]f,\text{ \ \ }\ell \lbrack u,v;\lambda
]:=\left(
\begin{array}{cc}
-\lambda u_{x} & -v_{x} \\
2\lambda ^{2} & \lambda u_{x}%
\end{array}%
\right) ,  \label{H24}
\end{equation}%
compatible with (\ref{H23}). Moreover, as a result of (\ref{H23}) and (\ref%
{H0d}), the general solution to (\ref{H24}) allows the following functional
representation:
\begin{eqnarray}
f_{1}(x,t) &=&\tilde{g}_{1}(u-tv,x-tu+vt^{2}/2),  \label{H24a} \\
f_{2}(x,t) &=&-t\lambda \tilde{g}_{1}(u-tv,x-tu+vt^{2}/2)+  \notag \\
&&+\tilde{g}_{2}(u-tv,x-tu+vt^{2}/2),  \notag
\end{eqnarray}%
where $\tilde{g}_{j}\in C^{\infty }(\mathbb{R}^{2};\mathbb{C}),j=\overline{%
1,2},$ are arbitrary smooth complex valued functions. Now combining together
the obtained relationships (\ref{H23}) and (\ref{H24}), we can formulate the
following proposition.

\begin{proposition}
The Riemann \ type hydrodynamical system (\ref{H0a}) is equivalent to a
completely integrable bi-Hamiltonian flow on the functional manifold $%
\mathcal{M},$ allowing the Lax type representation
\end{proposition}

\begin{equation}
\begin{array}{c}
f_{x}=\ell \lbrack u,v;\lambda ]f,\text{ \ \ }f_{t}=p(\ell )f,\text{ \ \ \ \
}p(\ell ):=-u\ell \lbrack u,v;\lambda ]+q(\lambda ), \\
\text{\ }\ell \lbrack u,v;\lambda ]:=\left(
\begin{array}{cc}
-\lambda u_{x} & -v_{x} \\
2\lambda ^{2} & \lambda u_{x}%
\end{array}%
\right) ,\text{ \ }q(\lambda ):=\left(
\begin{array}{cc}
0 & 0 \\
-\lambda & 0%
\end{array}%
\right) ,\text{\ } \\
p(\ell )=\left(
\begin{array}{cc}
\lambda u_{x}u & v_{x}u \\
-\lambda -2\lambda ^{2}u & -\lambda u_{x}u%
\end{array}%
\right) ,%
\end{array}
\label{H27}
\end{equation}%
where $f\in C^{\infty }(\mathbb{R}^{2};\mathbb{C}^{2})$ and $\lambda \in
\mathbb{C}$ is an arbitrary spectral parameter.\

\begin{remark}
\label{Rm_4}It is worth to mention here that equation (\ref{H23}) is
equivalent on the solution set of the Riemann type hydrodynamical system (%
\ref{H1-0}) to the alone equation
\begin{equation}
D_{t}^{2}f_{2}=0\Longleftrightarrow D_{t}f_{1}=0,D_{t}f_{2}=-\lambda f_{1,}
\label{H23a}
\end{equation}%
where vector $f\in C^{\infty }(\mathbb{R}^{2};\mathbb{C}^{2})$ $\ $satisfies
for all $\lambda \in \mathbb{C}$ the compatibility condition (\ref{H24}) and
whose general solution is represented in the functional form (\ref{H24a}).
\end{remark}

Concerning the set of conservation laws $\{H_{0}^{(1/2)},H_{1}^{(1/2)}\},$
constructed above, they can be extended to an infinite hierarchy $%
\{H_{j}^{(1/2)}\in D(\mathcal{M}):j\in \mathbb{Z}_{+}\cup \{-1\}\},$ where
\begin{equation}
H_{j}^{(1/2)}:=\int_{0}^{2\pi }\sigma _{2j+1}[u,v]dx,\text{ }  \label{H28}
\end{equation}%
and the affine generating function $\sigma (x;\lambda ):=\frac{d}{dx}\ln
f_{2}(x;\lambda )\simeq \sum_{j=\mathbb{Z}_{+}\cup \{-1\}}\sigma
_{j}[u,v]\lambda ^{-j}$ as $\lambda \rightarrow \infty $ satisfies the
following functional equation:%
\begin{equation}
(\sigma -\lambda u_{x})_{x}+\sigma ^{2}+\lambda ^{2}(2v_{x}-u_{x}^{2})=0.
\label{H29}
\end{equation}%
In addition, the gradient functional $\varphi (x;\lambda ):={grad}$ $\gamma
(x;\lambda )\in T^{\ast }(\mathcal{M}),$ where $\gamma (\lambda
):=\int_{0}^{2\pi }\sigma (x;\lambda )dx,$ satisfies for all $\lambda \in
\mathbb{C}$ the gradient relationship (\ref{H21a}).

\section{The generalized Riemann type hydrodynamical equation at N=3:
conservation laws, bi-Hamiltonian structure and Lax type representation}

Here we proceed to analyzing conservation laws and bi-Hamiltonian structure
of the generalized Riemann type equation (\ref{H0a}) at $N=3:$
\begin{equation}
\left.
\begin{array}{c}
u_{t}=v-uu_{x} \\
v_{t}=z-uv_{x} \\
z_{t}=-uz_{x}%
\end{array}%
\right\} :=K[u,v,z],  \label{H30}
\end{equation}%
where $K:\mathcal{M}\rightarrow T(\mathcal{M})$ is a suitable vector field
on the periodic functional manifold $\mathcal{M}:=C^{\infty }(\mathbb{R}%
/2\pi \mathbb{Z};\mathbb{R}^{3})$ and $t\in \mathbb{R}$ is an evolution
parameter. \ The system (\ref{H30}) proves also to possess infinite
hierarchies of polynomial conservation laws, being suspicious for complete
and Lax type integrability.

Namely, the following polynomial functionals are conserved with respect to
the flow (\ref{H30}):%
\begin{eqnarray}
H_{n}^{(1)} &:&=\int_{0}^{2\pi }dxz^{n}(vu_{x}-v_{x}u-\frac{n+2}{n+1}z),
\label{H31} \\
H^{(4)} &:&=\int_{0}^{2\pi
}dx[-7v_{x}v^{2}u+z(6zu+2v_{x}u^{2}-3v^{2}-4vuu_{x})],  \notag \\
H^{(5)} &:&=\int_{0}^{2\pi }dx(z^{2}u_{x}-2zvv_{x}),\text{ \ }%
H^{(6)}:=\int_{0}^{2\pi }dx(z_{z}v^{3}+3z^{2}v_{x}u+z^{3}),  \notag \\
H^{(7)} &:&=\int_{0}^{2\pi }dx(z_{x}v^{3}+3z^{2}vu_{x}-3z^{3}),  \notag \\
H^{(8)} &:&=\int_{0}^{2\pi
}dxz(6z^{2}u+3zv_{x}u^{2}-3zv^{2}-4zvu_{x}-2v_{x}v^{2}u+2v^{3}u_{x}),  \notag
\\
H^{(9)} &:&=\int_{0}^{2\pi
}dx[1001v_{x}v^{4}u+(1092z^{2}u^{2}+364zv_{x}u^{3}-  \notag \\
&&-1092zv^{2}u-728zvu_{x}u^{2}-364v_{x}v^{2}u^{2}+273v^{4}+728v^{3}u_{x}u]),
\notag \\
H_{n}^{(2)} &:&=\int_{0}^{2\pi }dxz_{x}vz^{n},\text{ \ \ }%
H_{n}^{(3)}:=\int_{0}^{2\pi }dxz_{x}(v^{2}-2zu)^{n},  \notag
\end{eqnarray}%
where $n\in \mathbb{Z}_{+}.$ In particular, as $n=1,2,...,$ from (\ref{H31})
one obtains that%
\begin{eqnarray}
H_{0}^{(2)} &:&=\int_{0}^{2\pi }dxz_{x}v,\text{ \ }H_{1}^{(2)}:=\int_{0}^{2%
\pi }dxz_{x}zv,...,  \label{H32} \\
H_{1}^{(3)} &:&=\int_{0}^{2\pi }dxz_{x}(v^{2}-2uz),\text{ }  \notag \\
H_{2}^{(3)} &:&=\int_{0}^{2\pi }dxz_{x}(v^{4}+4z^{2}u^{2}-4zv^{2}u),...,
\notag
\end{eqnarray}%
and so on.

Making use of the iterative property, similar to that, formulated above in
Proposition \ref{Pr_1}, one can construct the following hierarchy of
non-polynomial dispersive and dispersionless conservation laws:%
\begin{eqnarray}
H_{1}^{(1/4)} &=&\int_{0}^{2\pi
}dx(-2u_{xx}u_{x}z_{x}+u_{xx}v_{x}^{2}+2u_{x}^{2}z_{xx}-  \notag \\
&&-u_{x}v_{xx}v_{x}+3v_{xx}z_{x}-3v_{x}z_{xx})^{1/4},  \notag \\
H_{2}^{(1/3)} &=&\int_{0}^{2\pi }dx(-v_{xx}z_{x}+v_{x}z_{xx})^{1/3},\text{ }
\notag \\
H_{3}^{(1/3)} &=&\int_{0}^{2\pi }dx(v_{xx}u_{x}-v_{x}u_{xx}-z_{xx})^{1/3},
\notag \\
H_{1}^{(1/2)} &=&\int_{0}^{2\pi
}dx[-2vu_{x}z_{x}+v_{x}^{2}+z(-u_{x}v_{x}+3z_{x})]^{1/2},  \notag \\
H_{2}^{(1/2)} &=&\int_{0}^{2\pi
}dx(8u_{x}^{3}z_{x}-3u_{x}^{2}v_{x}^{2}-18u_{x}v_{x}z_{x}+6v_{x}^{3}+9z_{x})^{1/2},
\label{H31a} \\
H_{1}^{(1/5)} &=&\int_{0}^{2\pi
}dx(-2u_{xxx}u_{x}z_{x}+u_{xxx}v_{x}^{2}+6u_{xx}^{2}z_{x}-6u_{xx}u_{x}z_{xx}-
\notag \\
&&-3u_{xx}v_{xx}v_{x}+2u_{x}^{2}z_{xxx}-u_{x}v_{xxx}v_{x}+  \notag \\
&&+3u_{x}v_{xx}^{2}+3v_{xxx}z_{x}-3v_{x}z_{xxx})^{1/5},  \notag \\
H_{3}^{(1/3)} &=&\int_{0}^{2\pi
}dx[k_{1}u(-v_{xx}z_{x}+v_{x}z_{xx})+k_{1}v(u_{xx}z_{x}-u_{x}z_{xx})+  \notag
\\
&&+z(k_{2}u_{xx}v_{x}-k_{2}u_{x}v_{xx}+k_{1}z_{xx}+k_{2}z_{xx})+  \notag \\
&&+k_{3}(-3u_{x}v_{x}z_{x}+v_{x}^{3}+3z_{x}^{2})]^{1/3},  \notag
\end{eqnarray}%
where $k_{j}\in \mathbb{R},j=\overline{1,3},$ are arbitrary real numbers. \
Below we will attempt to generalize the crucial relationship (\ref{H23})
from Section 2 \ on the case of the Riemann type hydrodynamical system (\ref%
{H30}). Namely, we will assume, based on the Remark (\ref{Rm_1}), that there
exists its following linearization:%
\begin{equation}
D_{t}^{3}f_{3}(\lambda )=0,  \label{H31d}
\end{equation}%
modeling the starting generalized Riemann type hydrodynamical equation (\ref%
{H0a}) at $N=3,$ and where $f_{3}(\lambda )\in C^{\infty }(\mathbb{R}^{2};%
\mathbb{C})$ for all values of the parameter $\lambda \in \mathbb{C}.$ The
scalar equation (\ref{H31d}) can be easily rewritten as the system of three
linear equations
\begin{equation}
D_{t}f_{1}=0,\text{ \ \ }D_{t}f_{2}=\mu _{1}f_{1},\text{ \ }D_{t}f_{3}=\mu
_{2}f_{2}  \label{H31e}
\end{equation}%
where we have defined a vector $f:=(f_{1},f_{2},f_{2})^{\intercal }\in
C^{\infty }(\mathbb{R}^{2};\mathbb{C}^{3})$ and naturally introduced
constant numbers $\mu _{j}:=\mu _{j}(\lambda )\in \mathbb{C},$ $\ j=%
\overline{1,2}.$ \ It is easy to observe now that, owing to the former
result (\ref{H0d}), the system of equations (\ref{H31e}) allows the
following solution representation:%
\begin{eqnarray}
f_{1}(x,t) &=&\tilde{g}_{1}(u-tv+zt^{2}/2,v-zt,x-tu+vt^{2}/2-zt^{3}/6),
\notag \\
f_{2}(x,t) &=&t\mu _{1}\tilde{g}%
_{1}(u-tv+zt^{2}/2,v-zt,x-tu+vt^{2}/2-zt^{3}/6)+  \notag \\
&&+\tilde{g}_{2}(u-tv+zt^{2}/2,v-zt,x-tu+vt^{2}/2-zt^{3}/6),  \label{H31ee}
\\
f_{3}(x,t) &=&\mu _{1}\mu _{2}\frac{t^{2}}{2}\tilde{g}%
_{1}(u-tv+zt^{2}/2,v-zt,x-tu+vt^{2}/2-zt^{3}/6)+  \notag \\
&&+t\mu _{2}\tilde{g}_{2}(u-tv+zt^{2}/2,v-zt,x-tu+vt^{2}/2-zt^{3}/6)+  \notag
\\
&&+\tilde{g}_{3}(u-tv+zt^{2}/2,v-zt,x-tu+vt^{2}/2-zt^{3}/6),  \notag
\end{eqnarray}%
where $\tilde{g}_{j}\in C^{\infty }(\mathbb{R}^{3};\mathbb{C}),j=\overline{%
1,3},$ are arbitrary smooth complex valued functions. The system (\ref{H31e}%
) transforms into the equivalent \ vector equation
\begin{equation}
D_{t}f=q(\mu )f,\text{ \ }q(\lambda ):=\left(
\begin{array}{ccc}
0 & 0 & 0 \\
\mu _{1}(\lambda ) & 0 & 0 \\
0 & \mu _{2}(\lambda ) & 0%
\end{array}%
\right) ,  \label{H31f}
\end{equation}%
which should be compatible both with a suitably chosen equation for
derivative
\begin{equation}
f_{x}=\ell \lbrack u,v,z;\lambda ]f  \label{H31g}
\end{equation}%
with some matrix $\ell \lbrack u,v,z;\lambda ]\in SL(3;\mathbb{C}),$ defined
on the functional manifold $\mathcal{M},$ and with the Lie-Lax equation (\ref%
{H22}), rewritten as the following system of equations
\begin{equation}
D_{t}\varphi =\left(
\begin{array}{ccc}
0 & v_{x} & z_{x} \\
-1 & -u_{x} & 0 \\
0 & -1 & -u_{x}%
\end{array}%
\right) \varphi ,\text{ \ }D_{t}=\partial /\partial t+u\partial /\partial x,
\label{H22b}
\end{equation}%
where the vector $\varphi :=\varphi (x;\lambda )\in T^{\ast }(\mathcal{M})$
is considered as the one factorized by means of a solution $f\in C^{\infty }(%
\mathbb{R}^{2};\mathbb{C}^{3})$ to (\ref{H31g}), satisfying the identity (%
\ref{H21c}). Namely, it is assumed that the following quadratic
trace-relationship
\begin{equation}
\varphi =tr(\Phi \lbrack \lambda ;u,v,z]\text{ }f\otimes f^{\intercal })
\label{H22c}
\end{equation}%
holds for some vector valued matrix $\Phi \lbrack \lambda ;u,v,z]\in \mathbb{%
E}^{3}\otimes End$ $\mathbb{E}^{3},$ \ \ defined on the manifold $\mathcal{M}%
,$ where "$\otimes "$ means the standard tensor product of vectors from the
Euclidean space $\mathbb{E}^{3}.$ \ Making now use of the determining
expressions (\ref{H21c}), (\ref{H22c}) and (\ref{H31f}), one can find by
means of some slightly cumbersome but tedious calculations that $\mu
_{1}(\lambda )=\lambda ,$ $\mu _{2}(\lambda )=1,$ $\ \lambda \in \mathbb{C},$
and the matrix representation of the derivative (\ref{H31g})%
\begin{equation}
\ell \lbrack u,v,z;\lambda ]=\left(
\begin{array}{ccc}
\lambda u_{x} & -v_{x} & z_{x} \\
3\lambda ^{2} & -2\lambda u_{x} & \lambda v_{x} \\
6\lambda ^{2}r[u,v,z] & -3\lambda  & \lambda u_{x}%
\end{array}%
\right) ,  \label{H31i}
\end{equation}%
compatible with determining equation (\ref{H22b}), where a smooth mapping $r:%
\mathcal{M}\rightarrow \mathbb{R}$ satisfies the differential relationship
\begin{equation}
D_{t}r+u_{x}r=1.  \label{H31ii}
\end{equation}%
The latter possesses a wide set $\mathcal{R}$ of different solutions amongst
which there are the following:%
\begin{eqnarray}
r &\in &\mathcal{R}:=\{[(xv-u^{2}/2)/z]_{x},(v_{x}-u_{x}^{2}/6)z_{x}^{-1},%
\frac{u_{x}^{3}/3-u_{x}v_{x}+3z_{x}/2}{2u_{x}z_{x}-v_{x}^{2}},  \label{W6.12}
\\
&&(v_{x}v^{3}/6-u_{x}v^{2}z/2+uz_{x}(uz-v^{2})/6+vz^{2})z^{-3}\}.  \notag
\end{eqnarray}%
Note here that only the third element from the set (\ref{W6.12}) allows the
reduction $z=0$ to the case $N=2.$ Thus, the resulting Lax type
representation for the Riemann type dynamical system (\ref{H30}) ensues in
the form:
\begin{equation}
\begin{array}{c}
f_{x}=\ell \lbrack u,v,z;\lambda ]{f},\text{ \ \ }f_{t}=p(\ell )f,\text{ \ \
}p(\ell ):=-u\ell \lbrack u,v,z;\lambda ]+q(\lambda ), \\
\\
\\
\ell \lbrack u,v,z;\lambda ]=\left(
\begin{array}{ccc}
\lambda u_{x} & -v_{x} & z_{x} \\
3\lambda ^{2} & -2\lambda u_{x} & \lambda v_{x} \\
6\lambda ^{2}r[u,v,z] & -3\lambda  & \lambda u_{x}%
\end{array}%
\right) ,\text{ \ }q(\lambda ):=\left(
\begin{array}{ccc}
0 & 0 & 0 \\
\lambda  & 0 & 0 \\
0 & 1 & 0%
\end{array}%
\right) , \\
\\
\\
\text{ \ }p(\ell )=\left(
\begin{array}{ccc}
-\lambda uu_{x} & uv_{x} & -uz_{x} \\
-3u\lambda ^{2}+\lambda  & 2\lambda uu_{x} & -\lambda uv_{x} \\
-6\lambda ^{2}r[u,v,z]u & 1+3u\lambda  & -\lambda uu_{x}%
\end{array}%
\right) ,%
\end{array}
\label{H31j}
\end{equation}%
where $f\in C^{\infty }(\mathbb{R}^{2};\mathbb{C}^{3})$ and \ $\lambda \in
\mathbb{C}$ is a spectral parameter.

The next problem, which is of great interest, consists in proving that the
generalized hydrodynamical system (\ref{H30}) is a completely integrable
bi-Hamiltonian flow on the periodic functional manifold $\mathcal{M},$ as it
was proved above for the system (\ref{H1-0}).

That dynamical system (\ref{H30}) is bi-Hamiltonian that follows easily as a
simple corollary from the fact that it possesses the Lax type representation
(\ref{H31j}) and from the general Lie-algebraic integrability theory \cite%
{Fa,PM,Bl}. Taking into account that dynamical system (\ref{H30}) possesses
many (at least 4) Lax type representations, one derives that it possesses
many (at least 4) different pairs of compatible co-symplectic structures,
every of which generates its \ own infinite hierarchy of commuting to each
other conservation laws. Moreover, the involution of \ conservation laws
belonging to different hierarchies fails owing to their non-compatibility.
As finding of these \ structures is adjoint with cumbersome enough
analytical calculations, we present below only a one pair of related
co-symplectic structures, making use of the standard properties of determining
them  Lie-Lax equation (\ref{H10}). \ \

To tackle with the\ related task of retrieving the Hamiltonian structure of
the dynamical system (\ref{H30}), it is enough, as in Section 2, to
construct \cite{PM,HPP} exact non-symmetric solutions to the Lie-Lax equation%
\begin{equation}
\psi _{t}+K^{\prime ,\ast }\psi ={grad}\text{ }\mathcal{L},\text{ \ \ }\psi
^{\prime }\neq \psi ^{\prime ,\ast },  \label{H33}
\end{equation}%
for some functional $\mathcal{L}\in D(\mathcal{M}),$ where $\psi \in T^{\ast
}(\mathcal{M})$ is, in general, a quasi-local vector, such that the system (%
\ref{H30}) allows the following Hamiltonian representation:%
\begin{equation}
\begin{array}{c}
K[u,v,z]=-\eta \text{ }{grad}\text{ }H[u,v,z],\text{ } \\
H_{\eta }=(\psi ,K)-\mathcal{L},\text{ \ \ }\eta ^{-1}=\psi ^{^{\prime
}}-\psi ^{\prime ,\ast }.%
\end{array}
\label{H34}
\end{equation}%
As a test solution to (\ref{H33}) one can take the one
\begin{equation*}
\psi =(u_{x}/2,0,-z_{x}^{-1}u_{x}^{2}/2+z_{x}^{-1}v_{x})^{\intercal },\text{
\ \ \ }\mathcal{L}=\frac{1}{2}\int_{0}^{2\pi }(2z+vu_{x})dx,
\end{equation*}%
which gives rise to the following co-implectic operator:
\begin{equation}
\eta ^{-1}:=\psi ^{\prime }-\psi ^{\prime ,\ast }=\left(
\begin{array}{ccc}
\partial & 0 & -\partial u_{x}z_{x}^{-1} \\
0 & 0 & \partial z_{x} \\
-u_{x}z_{x}^{-1}\partial & z_{x}\partial &
\begin{array}{c}
\frac{1}{2}(u_{x}^{2}z_{x}^{-2}\partial +\partial u_{x}^{2}z_{x}^{-2})- \\
-(v_{x}z_{x}^{-2}\partial +\partial v_{x}z_{x}^{-2})%
\end{array}%
\end{array}%
\right) .  \label{H35}
\end{equation}%
This expression is not strictly invertible, as its kernel possesses the
translation vector field $d/dx:\mathcal{M}\rightarrow T(\mathcal{M})$ with
components $(u_{x},v_{x},z_{x})^{\intercal }\in T(\mathcal{M}),$ that is $%
\eta ^{-1}(u_{x},v_{x},z_{x})^{\intercal }=0.$

Nonetheless, upon formal inverting the operator expression (\ref{H35}), we
obtain by means of simple enough, but slightly cumbersome, direct
calculations, that the Hamiltonian function equals
\begin{equation}
H_{\eta }:=\int_{0}^{2\pi }dx(u_{x}v-z).  \label{H36}
\end{equation}%
and the implectic $\eta $-operator looks as

\begin{equation}
\eta :=\left(
\begin{array}{ccc}
\partial ^{-1} & u_{x}\partial ^{-1} & 0 \\
\partial ^{-1}u_{x} & v_{x}\partial ^{-1}+\partial ^{-1}v_{x} & \partial
^{-1}z_{x} \\
0 & z_{x}\partial ^{-1} & 0%
\end{array}%
\right) .  \label{H37}
\end{equation}

The same way, representing the Hamiltonian function (\ref{H36}) in the
scalar form%
\begin{equation}
H_{\eta }=(\psi ,(u_{x},v_{x},z_{x})^{\intercal }),\text{ \ \ }\psi =\frac{1%
}{2}(-v,\text{ }u+...,-\frac{1}{\sqrt{z}}\partial ^{-1}\sqrt{z})^{\intercal
},  \label{H38}
\end{equation}%
one can construct a second implectic (co-symplectic) operator $\vartheta
:T^{\ast }(\mathcal{M})\rightarrow T(\mathcal{M}),$ looking \ up to $O(\mu
^{2})$ terms, as follows:
\begin{equation}
\vartheta =\left(
\begin{array}{ccc}
\begin{array}{c}
\mu (\frac{(u^{(1)})^{2}}{z^{(1)}}\partial +\partial \frac{(u^{(1)})^{2}}{%
z^{(1)}}) \\
\\
\end{array}
&
\begin{array}{c}
1+\frac{2\mu }{3}(\frac{u^{(1)}v^{(1)}}{z^{(1)}}\partial +2\partial \frac{%
u^{(1)}v^{(1)}}{z^{(1)}}) \\
\\
\end{array}
&
\begin{array}{c}
\frac{2\mu }{3}(\partial \frac{(v^{(1)})^{2}}{z^{(1)}}+\partial u^{(1)}) \\
\\
\end{array}
\\
-1+\frac{2\mu }{3}(\partial \frac{u^{(1)}v^{(1)}}{z^{(1)}}+2\frac{%
u^{(1)}v^{(1)}}{z^{(1)}}\partial ) &
\begin{array}{c}
\frac{2\mu }{3}(\frac{(v^{(1)})^{2}}{z^{(1)}}\partial +\partial \frac{%
(v^{(1)})^{2}}{z^{(1)}})+ \\
\\
+\frac{2\mu }{3}(u^{(1)}\partial +\partial u^{(1)})%
\end{array}
& 2\mu \partial v^{(1)} \\
\begin{array}{c}
\\
\frac{2\mu }{3}(\frac{(v^{(1)})^{2}}{z^{(1)}}\partial +u^{(1)}\partial )%
\end{array}
&
\begin{array}{c}
\\
2\mu v^{(1)}\partial%
\end{array}
&
\begin{array}{c}
\\
\mu (\partial z^{(1)}+z^{(1)}\partial )%
\end{array}%
\end{array}%
\right) +O(\mu ^{2}),  \label{H39}
\end{equation}%
where we put, by definition, $\vartheta ^{-1}:=(\psi ^{\prime }-\psi
^{\prime ,\ast }),$ $u:=\mu u^{(1)},v:=\mu v^{(1)},z:=\mu z^{(1)}$ as $\mu
\rightarrow 0,$ and whose exact form needs some additional simple but
cumbersome calculations, which will be presented in a work under preparation.

The operator (\ref{H39}) satisfies the Hamiltonian vector field condition:
\begin{equation}
(u_{x},v_{x},z_{x})^{\intercal }=-\vartheta \text{ }{grad}\text{ }H_{\eta },
\label{H40}
\end{equation}%
following easily from (\ref{H38}).

The results obtained above can be formulated as the following proposition.

\begin{proposition}
The Riemann \ type hydrodynamical system (\ref{H0a}) at $N=3$ is equivalent
to a completely integrable bi-Hamiltonian flow on the functional manifold $%
\mathcal{M},$ allowing the Lax type representation (\ref{H31j}) and the
compatible pair of \ co-symplectic structures (\ref{H37}) and (\ref{H39}).
\end{proposition}

The infinite hierarchy of conservation laws like (\ref{H31a}) \ and related
recurrent relationships can be regularly reconstructed, if to compute the
asymptotical solutions to the following Lie-Lax equation:%
\begin{eqnarray}
L_{\tilde{K}}\tilde{\varphi} &=&\tilde{\varphi}_{\tau }+\tilde{K}^{\prime
,\ast }\tilde{\varphi}=0,  \notag \\
\tilde{\varphi} &\simeq &\tilde{a}(x;\lambda )\exp \{\lambda ^{2}\tau
+\partial ^{-1}\tilde{\sigma}(x;\lambda )\},  \label{H31b}
\end{eqnarray}%
where, by definition, $\tilde{a}(x;\lambda )\simeq \sum_{j\in \mathbb{Z}_{+}}%
\tilde{a}_{j}[u,v,z]\lambda ^{-j},$ $\tilde{\sigma}(x;\lambda )\simeq
\sum_{j\in \mathbb{Z}_{+}\cup \{-1\}}\tilde{\sigma}_{j}[u,v,z]\lambda ^{-j}$
as $\lambda \rightarrow \infty ,$ and
\begin{eqnarray}
\frac{d}{d\tau }(u,v,z)^{\intercal } &:&=-3\eta \text{ }{grad}\text{ }%
H_{3}^{(1/3)}[u,v,z]=  \label{H31c} \\
&=&\left.
\begin{array}{c}
\begin{array}{c}
-(u_{x}^{2}h^{-2})_{x}+v_{x}^{-1}(v_{x}^{2}h^{-2})_{x} \\
-v_{x}u_{x}^{-1}(u_{x}^{2}h^{-2})_{x}+z_{x}^{-1}(z_{x}^{2}h^{-2})_{x} \\
-z_{x}u_{x}^{-1}(z_{x}^{2}h^{-2})_{x}%
\end{array}%
\end{array}%
\right\} :=\tilde{K}[u,v,z],  \notag \\
H_{3}^{(1/3)} &:&=\int_{0}^{2\pi }h[u,v,z]dx,\text{ \ \ }%
h[u,v,z]=(v_{xx}u_{x}-u_{xx}v_{x}-z_{xx})^{1/3},  \notag
\end{eqnarray}%
is a Hamiltonian vector field on the functional manifold $\mathcal{M}$ with
respect to a suitable evolution parameter $\tau \in \mathbb{R}.$\ \ Since
the vector fields (\ref{H31c}) and (\ref{H30}) are commuting to each other
on the whole manifold $\mathcal{M},$ the functionals $\tilde{H}%
_{j}^{(1/3)}:=\int_{0}^{2\pi }\tilde{\sigma}_{j}[u,v,z]dx,$ \ $j\in \mathbb{Z%
}_{+}\cup \{-1\},$ will be functionally independent conservation laws for
both these dynamical systems.

The Lax type integrability of the Riemann type hydrodynamical equation (\ref%
{H0a}) at $N=2$ and $N=3,$ stated above, \ allows one to speculate that it
is also integrable for arbitrary $N\in \mathbb{Z}_{+}.$

Concerning the \ evident difference between analytical properties of the
cases $N=2$ and $N=3$, we can easily observe that it is related with
structures of the corresponding Lax type operators (\ref{H24}) and (\ref%
{H31j}): in the first case the corresponding $r$-equation (\ref{H31ii}) is
trivial (that is empty), but in the second case it is already nontrivial,
allowing many different solutions. This situation generalizes, as we will
see below, to the case $N\geq 4,$ thereby explaining the appearing diversity
of related Lax type representations.

To support this hypothesis we will prove below that also at $N=4$ it is
equivalent to a Lax type integrable bi-Hamiltonian dynamical system on the
suitable smooth $2\pi $-periodic functional manifold $\mathcal{M}:=$ $%
C^{\infty }(\mathbb{R}/2\pi \mathbb{Z};\mathbb{R}^{4}\mathbb{)},$ possesses
infinite hierarchies of polynomial dispersionless and dispersive
non-polynomial conservation laws.

\section{The generalized Riemann type hydrodynamical equation at N=4:
conservation laws, bi-Hamiltonian structure and Lax type representation}

The Riemann type hydrodynamical equation (\ref{H0a}) at $N=4$ is equivalent
to the nonlinear dynamical system%
\begin{equation}
\left.
\begin{array}{c}
u_{t}=v-uu_{x} \\
v_{t}=w-uv_{x} \\
w_{t}=z-uw_{x} \\
z_{t}=-uz_{x}%
\end{array}%
\right\} :=K[u,v,w,z],  \label{H41}
\end{equation}%
where $K:\mathcal{M}\rightarrow T(\mathcal{M})$ is a suitable vector field
on the smooth $2\pi $-periodic functional manifold $\mathcal{M}:=$ $%
C^{\infty }(\mathbb{R}/2\pi \mathbb{Z};\mathbb{R}^{4}\mathbb{)}.$ To state
its Hamiltonian structure, we need to find an exact non-symmetric functional
solution $\psi \in T^{\ast }(\mathcal{M})$ to the Lie-Lax equation (\ref{H33}%
):%
\begin{equation}
\psi _{t}+K^{^{\prime ,\ast }}\psi =grad\text{ }\mathcal{L}  \label{H42}
\end{equation}%
for some smooth functional $\mathcal{L}\in D(\mathcal{M}),$ where
\begin{equation}
K^{^{\prime }}=\left(
\begin{array}{cccc}
-\partial u & 1 & 0 & 0 \\
-v_{x} & -u\partial & 1 & 0 \\
-w_{x} & 0 & -u\partial & 1 \\
-z_{x} & 0 & 0 & -u\partial%
\end{array}%
\right) ,\text{ \ }K^{^{\prime ,\ast }}=\left(
\begin{array}{cccc}
u\partial & -v_{x} & -w_{x} & -z_{x} \\
1 & \partial u & 0 & 0 \\
0 & 1 & \partial u & 0 \\
0 & 0 & 1 & \partial u%
\end{array}%
\right)  \label{H43}
\end{equation}%
are, respectively, the Frechet derivative of the mapping $K:\mathcal{M}%
\rightarrow T(\mathcal{M})$ and its conjugate. \ The small parameter method
\cite{PM}, applied to equation (\ref{H42}), gives rise to the following its
exact solution:%
\begin{eqnarray}
\psi &=&(-w_{x},v_{x}/2,0,-\frac{v_{x}^{2}}{2z_{x}}+\frac{u_{x}w_{x}}{z_{x}}%
)^{\intercal },  \label{H44} \\
\mathcal{L} &=&\int_{0}^{2\pi }(zu_{x}-vw_{x}/2)dx.  \notag
\end{eqnarray}%
As a result, we obtain right away from (\ref{H42}) that dynamical system (%
\ref{H41}) is a Hamiltonian system on the functional manifold $\mathcal{M},$
that is
\begin{equation}
K=-\vartheta \text{ }grad\text{ }H,  \label{H45}
\end{equation}%
where the Hamiltonian functional equals
\begin{equation}
H:=(\psi ,K)-\mathcal{L}=\int_{0}^{2\pi }(uz_{x}-vw_{x})dx  \label{H46}
\end{equation}%
and the co-implectic operator equals
\begin{equation}
\vartheta ^{-1}:=\psi ^{\prime }-\psi ^{\prime ,\ast }=\left(
\begin{array}{cccc}
0 & 0 & -\partial & \partial \frac{w_{x}}{z_{x}} \\
0 & -u\partial & 0 & -\partial \frac{v_{x}}{z_{x}} \\
-\partial & 0 & 0 & \partial \frac{u_{x}}{z_{x}} \\
\frac{w_{x}}{z_{x}}\partial & -\frac{v_{x}}{z_{x}}\partial & \frac{u_{x}}{%
z_{x}}\partial &
\begin{array}{c}
\frac{1}{2}[z_{x}^{-2}(v_{x}^{2}-2u_{x}w_{x})\partial + \\
+\partial (v_{x}^{2}-2u_{x}w_{x})z_{x}^{-2}]%
\end{array}%
\end{array}%
\right) .  \label{H47}
\end{equation}%
The latter is degenerate: \ the relationship $\vartheta
^{-1}(u_{x},v_{x},w_{x},z_{x})^{\intercal }=0$ holds exactly on the whole
manifold $\mathcal{M},$ but the inverse to (\ref{H47}) exists and can be
calculated analytically.

To state the Lax type integrability of Hamiltonian system (\ref{H41}) we
will apply to it, as in Section 3 above, the standard gradient-holonomic
scheme of \cite{PM,HPP} and find the following its linearization:%
\begin{equation}
D_{t}^{4}f_{4}(\lambda )=0,  \label{H48}
\end{equation}%
where $f_{4}(\lambda )\in $ $C^{\infty }(\mathbb{R}^{2};\mathbb{C})$ for all
$\lambda \in \mathbb{C}.$ Having rewritten (\ref{H48}) in the form of the
linear system
\begin{equation}
D_{t}f=q(\lambda )f,\text{ \ }q(\lambda ):=\left(
\begin{array}{cccc}
0 & 0 & 0 & 0 \\
\lambda & 0 & 0 & 0 \\
0 & \lambda & 0 & 0 \\
0 & 0 & \lambda & 0%
\end{array}%
\right)  \label{H49}
\end{equation}%
with $\lambda \in \mathbb{C}$ \ \ being an arbitrary constant, $\ \ $for the
vector $f\in $ $C^{\infty }(\mathbb{R}^{2};\mathbb{C}^{4})$ one obtains
easily, owing to the relationships (\ref{H0d}), the following functional
representation:
\begin{eqnarray}
f_{1}(x,t) &=&\tilde{g}_{1}(u-tv+wt^{2}/2-xt^{3}/3!,v-wt+zt^{2}/2,w-zt,
\notag \\
&&x-tu+vt^{2}/2-wt^{3}/3!+zt^{4}/4!),  \notag \\
f_{2}(x,t) &=&t\lambda \tilde{g}%
_{1}(u-tv+wt^{2}/2-xt^{3}/3!,v-wt+zt^{2}/2,w-zt,  \notag \\
&&x-tu+vt^{2}/2-wt^{3}/3!+zt^{4}/4!)+  \notag \\
&&+\tilde{g}_{2}(u-tv+wt^{2}/2-xt^{3}/3!,v-wt+zt^{2}/2,w-zt,  \notag \\
&&x-tu+vt^{2}/2-wt^{3}/3!+zt^{4}/4!),  \notag \\
f_{3}(x,t) &=&\lambda ^{2}\frac{t^{2}}{2}\tilde{g}%
_{1}(u-tv+wt^{2}/2-xt^{3}/3!,v-wt+zt^{2}/2,w-zt,  \notag \\
&&x-tu+vt^{2}/2-wt^{3}/3!+zt^{4}/4!)+  \notag \\
&&+t\lambda \tilde{g}_{2}(u-tv+wt^{2}/2-xt^{3}/3!,v-wt+zt^{2}/2,w-zt,  \notag
\\
&&x-tu+vt^{2}/2-wt^{3}/3!+zt^{4}/4!)+  \notag \\
&&+\tilde{g}_{3}(u-tv+wt^{2}/2-xt^{3}/3!,v-wt+zt^{2}/2,w-zt,  \label{H50} \\
&&x-tu+vt^{2}/2-wt^{3}/3!+zt^{4}/4!),  \notag \\
f_{4}(x,t) &=&\lambda ^{3}\frac{t^{3}}{3!}\tilde{g}%
_{1}(u-tv+wt^{2}/2-xt^{3}/3!,v-wt+zt^{2}/2,w-zt,  \notag \\
&&x-tu+vt^{2}/2-wt^{3}/3!+zt^{4}/4!)+  \notag \\
&&+\lambda ^{2}\frac{t^{2}}{2}\tilde{g}%
_{2}(u-tv+wt^{2}/2-xt^{3}/3!,v-wt+zt^{2}/2,w-zt,  \notag \\
&&x-tu+vt^{2}/2-wt^{3}/3!+zt^{4}/4!)+  \notag \\
&&+t\lambda \tilde{g}_{3}(u-tv+wt^{2}/2-xt^{3}/3!,v-wt+zt^{2}/2,w-zt,  \notag
\\
&&x-tu+vt^{2}/2-wt^{3}/3!+zt^{4}/4!)+  \notag \\
&&+\tilde{g}_{4}(u-tv+wt^{2}/2-xt^{3}/3!,v-wt+zt^{2}/2,w-zt,  \notag \\
&&x-tu+vt^{2}/2-wt^{3}/3!+zt^{4}/4!),  \notag
\end{eqnarray}%
where $\tilde{g}_{j}\in C^{\infty }(\mathbb{R}^{4};\mathbb{C}),j=\overline{%
1,4},$ are arbitrary smooth complex valued functions.

Based now on the expressions (\ref{H49}) and (\ref{H50}), one can construct
the related Lax type representation for dynamical system (\ref{H41}) in the
following \ compatible form:
\begin{equation}
\begin{array}{c}
f_{x}=\ell \lbrack u,v,z,w;\lambda ]{f},\text{ \ \ }f_{t}=p(\ell )f,\text{ \
\ }p(\ell ):=-u\ell \lbrack u,v,w,z;\lambda ]+q(\lambda ),%
\end{array}
\label{H51}
\end{equation}%
where
\begin{equation*}
\ell \lbrack u,v,w,z;\lambda ]:=\left(
\begin{array}{cccc}
-\lambda ^{3}u_{x} & \lambda ^{2}v_{x} & -\lambda w_{x} & z_{x} \\
-4\lambda ^{4} & 3\lambda ^{3}u_{x} & -2\lambda ^{2}v_{x} & \lambda w_{x} \\
-10\lambda ^{5}r_{1} & 6\lambda ^{4} & -3\lambda ^{3}u_{x} & \lambda
^{2}v_{x} \\
-20\lambda ^{6}r_{2} & 10\lambda ^{5}r_{1} & -4\lambda ^{4} & \lambda
^{3}u_{x}%
\end{array}%
\right) ,\text{ \ }q(\lambda ):=\left(
\begin{array}{cccc}
0 & 0 & 0 & 0 \\
\lambda & 0 & 0 & 0 \\
0 & \lambda & 0 & 0 \\
0 & 0 & \lambda & 0%
\end{array}%
\right) ,
\end{equation*}%
\begin{equation}
\text{ \ }p(\ell )=\left(
\begin{array}{cccc}
\lambda uu_{x} & -\lambda ^{2}uv_{x} & \lambda uw_{x} & -uz_{x} \\
\lambda +4\lambda ^{4}u & -3\lambda ^{3}uu_{x} & 2\lambda ^{2}uv_{x} &
-\lambda uw_{x} \\
10\lambda ^{5}ur_{1} & \lambda -6\lambda ^{4}u & 3\lambda ^{3}uu_{x} &
-\lambda ^{2}uv_{x} \\
20\lambda ^{6}ur_{2} & -10\lambda ^{5}ur_{1} & \lambda +4\lambda ^{4}u &
-\lambda ^{3}uu_{x}%
\end{array}%
\right) ,  \label{H51a}
\end{equation}%
the mappings $r_{j}:\mathcal{M}\rightarrow \mathbb{R},j=\overline{1,2},$
satisfy the functional-differential equations
\begin{equation}
D_{t}r_{1}+r_{1}D_{x}u=1,\text{ \ \ \ \ \ \ }D_{t}r_{2}+r_{2}D_{x}u=r_{1},
\label{A16}
\end{equation}%
similar to (\ref{H31ii}), considered already above, thereby being a Lax type
integrable dynamical system on the functional manifold $\mathcal{M}.$

The equations (\ref{A16}), as it is easy to demonstrate \cite{BGPPP,PAPP} by
means of standard differential-algebraic methods, possess a lot of different
solutions, amongst which there are functional expressions:
\begin{eqnarray}
r_{1} &=&D_{x}(\frac{uw^{2}}{2z^{2}}-\frac{vw^{3}}{3z^{3}}+\frac{vw^{4}}{%
24z^{4}}+\frac{7w^{5}}{120z^{4}}-\frac{w^{6}}{144z^{5}}),  \label{A17} \\
r_{2} &=&D_{x}(\frac{uw^{3}}{3z^{3}}-\frac{vw^{4}}{6z^{4}}+\frac{3w^{6}}{%
80z^{5}}+\frac{vw^{5}}{120z^{5}}-\frac{w^{7}}{420z^{6}}).  \notag
\end{eqnarray}

Owing to the existence of the Lax type representation (\ref{H51}), (\ref%
{H51a}) and the related gradient like relationship (\ref{H21a}), we can
easily derive that the Hamiltonian system (\ref{H41}) is also simultaneously
a bi-Hamiltonian flow on the functional manifold $\mathcal{M}.$ And as it
was mentioned before, it possesses many bi-Hamiltonian structures, depending
on a chosen solution to the corresponding functional-differential equations (%
\ref{A16}).

In addition, we can construct, making use of the results above and the
approach of Section 1, the infinite hierarchies of related conservation laws
for (\ref{H41}), both dispersionless polynomial and dispersive
non-polynomial ones:

\bigskip

a) polynomial conservation laws:%
\begin{eqnarray}
H^{(9)} &=&\int_{0}^{2\pi }dx(vw_{x}-uz_{x}),\ \text{\ \ \ \ }%
H^{(19)}=\int_{0}^{2\pi }dxz_{x}(w^{2}-2vz),  \label{H52} \\
H^{(16)} &=&\int_{0}^{2\pi }dx(3u_{x}z^{2}+4w_{x}vz+2z_{x}vw),\text{ \ }%
H^{(20)}=\int_{0}^{2\pi }dx(z_{x}w-zw_{x}),  \notag \\
H^{(17)} &=&\int_{0}^{2\pi }dx[3u_{x}z(3uz+2vw)-6v_{x}z(uw+v^{2})+  \notag \\
&&+6w_{x}(uvz+2uw^{2}-v^{2}w)+z(w^{2}-2vz)],  \notag \\
H^{(18)} &=&\int_{0}^{2\pi
}dx[k_{1}(z_{x}(2uw-v^{2})+z^{2})+k_{2}((2w_{x}(uz-vw)+2z_{x}(v^{2}-uw))],
\notag \\
&&  \notag
\end{eqnarray}%
b) non-polynomial conservation laws:
\begin{eqnarray}
&&  \label{H53} \\
H^{(10)} &=&\int_{0}^{2\pi }dx(w_{x}^{2}-2v_{x}z_{x})^{1/2},  \notag \\
&&  \notag \\
H^{(11)} &=&\int_{0}^{2\pi }dx\Big (%
u_{xx}z_{x}-u_{x}z_{xx}+v_{x}w_{xx}-v_{xx}w_{xx}\Big )^{\frac{1}{3}},  \notag
\\
&&  \notag \\
H^{(12)} &=&\int_{0}^{2\pi }dx\Big (%
9u_{x}^{2}z_{x}-6u_{x}v_{x}w_{x}+2v_{x}^{3}-12v_{x}z_{x}+6w_{x}^{2}\Big )^{%
\frac{1}{3}},  \notag \\
&&  \notag \\
\text{\ }H^{(13)} &=&\int_{0}^{2\pi }dx\Big (%
u(2v_{x}z_{x}-w_{x}^{2})+v(v_{x}w_{x}-3u_{x}z_{x})+  \notag \\
&&+w(u_{x}w_{x}-v_{x}^{2}+2z_{x})+z(u_{x}v_{x}-2w_{x})\Big )^{\frac{1}{3}}],
\notag \\
&&  \notag \\
H^{(14)} &=&k_{1}H_{1}^{(14)}+k_{2}H_{2}^{(14)}+H_{3}^{(14)},\text{ \ \ \ }%
H^{(15)}=k_{1}H_{1}^{(15)}+k_{2}H_{2}^{(15)}+H_{3}^{(15)},  \notag
\end{eqnarray}%
where%
\begin{eqnarray*}
H_{1}^{(14)} &=&\int_{0}^{2\pi }\Big(u_{xx}(2v_{x}z_{z}-w_{x}^{2}\big)+v_{xx}%
\big(v_{x}w_{x}-3u_{x}z_{x}\big )+ \\
&&+w_{xx}\big(u_{x}w_{x}-v_{x}^{2}+2z_{x}\big)+z_{xx}\big(u_{x}v_{x}-2w_{x}%
\big)\Big)^{\frac{1}{4}}, \\
&& \\
H_{2}^{(14)} &=&\int_{0}^{2\pi }dx\Big (z_{x}w_{xx}-z_{xx}w_{x}\Big )^{\frac{%
1}{3}}, \\
&& \\
H_{3}^{(14)} &=&\int_{0}^{2\pi }dx[k_{1}\big(%
v(2v_{x}z_{x}-w_{x}^{2})+z(4z_{x}-u_{x}w_{x})+w(v_{x}w_{x}-3u_{x}z_{x})\big)+
\\
&&+k_{2}z\big(2z_{x}+v_{x}^{2}-u_{x}w_{x}\big )]^{\frac{1}{2}}, \\
&&
\end{eqnarray*}

\begin{eqnarray}
H_{1}^{(15)} &=&\int_{0}^{2\pi
}dx[u_{xxx}(2v_{x}z_{x}-w_{x}^{2})+v_{xxx}(v_{x}w_{w}-3u_{x}z_{x})+  \notag
\\
&&+z_{xxx}(u_{x}v_{x}-2w_{x})+w_{xxx}(u_{x}w_{x}-v_{x}^{2}+2z_{x})+  \notag
\\
&&+3u_{xx}(v_{xx}z_{x}-3v_{x}z_{xx}+w_{xx}w_{x})+3v_{xx}(2u_{x}z_{xx}-
\label{H53a} \\
&&-v_{xx}w_{x}+v_{x}w_{xx})-3w_{xx}^{2}u_{x}]^{\frac{1}{5}},  \notag \\
H_{2}^{(15)} &=&\int_{0}^{2\pi }dx\Big (%
4u_{x}^{2}w_{x}^{2}-4u_{x}v_{x}^{2}w_{x}-8u_{x}z_{x}w_{x}+v_{x}^{4}++4v_{x}^{2}z_{x}+4z_{x}^{2}%
\Big )^{\frac{1}{4}},  \notag \\
&&  \notag \\
H_{3}^{(15)} &=&\int_{0}^{2\pi }dx\left\{
k_{3}[u(z_{x}w_{xx}-z_{xx}w_{x})\right. +  \notag \\
&&+v(v_{x}z_{xx}-v_{xx}z_{x})+zz_{xx}+w(u_{xx}z_{x}-u_{x}z_{xx})]+  \notag \\
&&+k_{4}[z(u_{xx}w_{x}-u_{x}w_{xx}+2z_{xx})+w(u_{xx}z_{x}-u_{x}z_{xx}-v_{xx}w_{x}+v_{x}w_{xx})]+
\notag \\
&&+\left. k_{5}z_{x}(v_{x}^{2}-2u_{x}w_{x}+2z_{x})\right\} ^{\frac{1}{3}},
\notag
\end{eqnarray}%
and $k_{j}\in \mathbb{R},$ $\ j=\overline{1,5},$ are arbitrary constants. We
observe also that the Hamiltonian functional (\ref{H46}) coincides exactly
up the sign with the polynomial conservation law $H^{(9)}\in D(\mathcal{M}).$

\begin{remark}
It is worth here to remark \cite{BGPPP} that the generalized Riemann type
hydrodynamical equation (\ref{H0a}) can be once more naturally generalized
to the following also integrable Riemann type equation
\begin{equation}
D_{t}^{N}u=0,\text{ \ }D_{t}:=\partial /\partial x+a(\hat{u})\partial
/\partial t,  \label{H54}
\end{equation}%
where $N\in \mathbb{Z}_{+}$ and $a\in C^{\infty }(\mathcal{M};\mathbb{R})$
is an arbitrary smooth mapping. The corresponding to (\ref{H54}) nonlinear
dynamical system
\begin{eqnarray}
u_{t}^{(0)} &=&u^{(1)}-a(\hat{u})u_{x}^{(1)},  \notag \\
u_{t}^{(1)} &=&u^{(2)}-a(\hat{u})u_{x}^{(2)},  \notag \\
&&......  \label{H55} \\
u_{t}^{(N-2)} &=&u^{(N-1)}-a(\hat{u})u_{x}^{(N-2)}  \notag \\
u_{t}^{(N-1)} &=&-a(\hat{u})u_{x}^{(N-1)}.  \notag
\end{eqnarray}%
will be also a bi-Hamiltonian Lax type integrable dynamical system on the
phase space $\mathcal{M}.$
\end{remark}

Thereby, the calculations above \ ensue the formulation of the following
proposition.

\begin{proposition}
The Riemann \ type hydrodynamical system (\ref{H0a}) at $N=4$ is equivalent
to a completely integrable bi-Hamiltonian flow on the functional manifold $%
\mathcal{M},$ allowing the Lax type representation (\ref{H51}) and whose
co-implectic structure is given by expression (\ref{H47}).
\end{proposition}

Concerning the general case $N\in \mathbb{Z}_{+},$ \ applying successively
either the symplectic approach devised above or the differential-algebraic
method devised in \cite{BGPPP,PAPP}, one can also obtain for both the
Riemann type hydrodynamical system (\ref{H0a}) and (\ref{H54}) the infinite
hierarchies of \ dispersive and dispersionless conservation laws,
co-symplectic structures and related Lax type representations, what is a
topic of the next work under preparation.

\section{Conclusion}

As follows from the results obtained in this work, the generalized Riemann
type hydrodynamical equation (\ref{H0a}) possesses many infinite hierarchies
of conservation laws, both dispersive non-polynomial and dispersionless
polynomial. This fact can be easily explained by the fact that the
corresponding dynamical system (\ref{H0b}) allows many, plausibly, infinite
set of algebraically independent compatible implectic structures, which
generate via the standard gradient like relationship (\ref{H20}) the related
infinite hierarchies of conservation laws, and as a by-product, infinite
hierarchies of the associated Lax type representations. Such a situation
within the theory of Lax type integrable nonlinear dynamical systems meets,
virtually, for the first time and may appear to be interesting from
different point of view, as well as theoretical and practical. Keeping in
mind these and some other important aspects of the generalized Riemann type
hydrodynamical equation (\ref{H0a}), we consider that they deserve
additional thorough investigation in the future.

\bigskip

\section{\protect\bigskip Acknowledgements}

Authors are sincerely appreciated to Profs. F. Calogero, M. Pavlov, M. B\l %
aszak, Z. Peradzy\'{n}ski, J. S\l awianowski, N. Bogolubov (jr.) and D.
Blackmore for useful discussions of the results obtained. The warm thanks go
to our colleagues Dr. J. Golenia and \ Dr. P. Holod for instrumental help in
editing the manuscript. The last but not least thanks go to Referees who
generously mentioned some important points related with the integrability
problem treated in the work.

\end{document}